# Verification of Embedded Memory Systems using Efficient Memory Modeling

Malay K Ganai, Aarti Gupta, Pranav Ashar
*{malay | agupta | ashar }@nec-labs.com*
NEC Laboratories America, Princeton, NJ USA 08540

## Abstract

*We describe verification techniques for embedded memory systems using efficient memory modeling (EMM), without explicitly modeling each memory bit. We extend our previously proposed approach of EMM in Bounded Model Checking (BMC) for a single read/write port single memory system, to more commonly occurring systems with multiple memories, having multiple read and write ports. More importantly, we augment such EMM to providing correctness proofs, in addition to finding real bugs as before. The novelties of our verification approach are in a) combining EMM with proof-based abstraction that preserves the correctness of a property up to a certain analysis depth of SAT-based BMC, and b) modeling arbitrary initial memory state precisely and thereby, providing inductive proofs using SAT-based BMC for embedded memory systems. Similar to the previous approach, we construct a verification model by eliminating memory arrays, but retaining the memory interface signals with their control logic and adding constraints on those signals at every analysis depth to preserve the data forwarding semantics. The size of these EMM constraints depends quadratically on the number of memory accesses and the number of read and write ports; and linearly on the address and data widths and the number of memories. We show the effectiveness of our approach on several industry designs and software programs.*

## 1. Introduction

According to the Semiconductor Industry Association roadmap prediction, embedded memories will comprise more than 70% of the SoC by 2005. These embedded memories on SoC support diverse code and data requirements arising from ever increasing demand for data throughput in applications ranging from cellular phones, smart cards and digital cameras. In the past, there were efforts [1] to verify on-chip memory arrays using Symbolic Trajectory Evaluation [2]. However, these embedded memories dramatically increase both design and verification complexity due to an exponential increase in the state space with each additional memory bit. In particular, this state explosion adversely affects the practical application of formal verification techniques like model checking [3, 4] for functional verification of such large embedded memory systems.

To tame the verification complexity, several memory abstraction techniques, i.e., removing the memories partially or completely from the designs are often used in the industry. However, such techniques often produce spurious outcomes, adversely affecting overall verification efforts. In many refinement-based techniques [5-8], starting from a small abstract model of the concrete design, spurious counter-examples on the abstract model are used to refine the model iteratively. In practice, several iterations are needed before a property can be proved correct or a real counter-example can be found. Note that after every iterative refinement step, the model size increases, making it increasingly difficult to verify. In contrast, abstraction-based approaches [9, 10] use proof-based abstraction (PBA) techniques on a concrete design. As shown in [10], iterative abstraction can be used to apply such techniques on progressively more abstract models, thereby leading to significant reduction in model size. However, since these approaches use the concrete model to start with, it may not be feasible to apply them on designs with large memories. In general, both these refinement and abstraction based approaches are not geared towards exploiting the memory semantics.

Memory abstractions that preserve the memory semantics – data read from a memory location is the same as the most recent data written at the same location – have been employed in various verification efforts in the past. Burch *et al.* introduced the interpreted *read* and *write* operations in their logic of equality with un-interpreted functions [11]. Such partial interpretation of memory has also been exploited in later derivative verification efforts [12-14]. Specifically, Velev *et al.* used this partial interpretation in a symbolic simulation engine to replace memory by a behavioral model that interacts with the rest of the circuit through a software interface that monitors the memory control signals [12]. Bryant *et al.* proposed [15] modeling of memory as a functional expression in the UCLID system for verifying safety properties.

SAT-based Bounded Model Checking (BMC) [16] enjoys several nice properties over BDD-based symbolic model checking [3, 4]; its performance is less sensitive to the problem sizes and it does not suffer from space explosion. To address the memory explosion problem, SAT-based distributed BMC has been proposed [17] in which the BMC problem is partitioned over a network of workstations. However, this technique is not geared towards verifying embedded memory systems. In our previous work, we have proposed an efficient memory modeling (EMM) technique [18] that augments SAT-based BMC to handle large embedded memories without explicitly modeling each memory bit. We showed that EMM approach allows deeper BMC search in finding real bugs in comparison to explicit memory models. Moreover, our approach captures the exclusivity of a matching read and writes pair explicitly, reducing the overall SAT solve time. However, there are two main drawbacks to this previous work. First, the memory system considered was fairly simplistic, with a single memory having a single read/write port. In modern designs, it is quite common to have a large number of diverse memories, each with multiple read and write ports. Second, the approach was geared towards falsification i.e., finding real bugs, and not towards proving correctness of the specified property.

In this work, we extend our previous approach [18] of EMM in SAT-based BMC to the more commonly occurring embedded memory systems, with multiple memories having multiple read and write ports. More importantly, we augment the EMM techniques to providing correctness proofs in addition to finding real bugs. The novelties of our verification approach are in a) combining EMM with PBA that preserves the correctness of a property up to a certain analysis depth of SAT-based BMC, and b) modeling arbitrary initial memory state precisely and thereby, providing inductive proofs using SAT-based BMC for embedded memory systems. Similar to the previous approach [18], we construct a verification model by eliminating memory arrays, but retaining the memory interface signals with their control logic and adding constraints on those signals at every analysis depth to preserve the memory data forwarding semantics. The size of these memory-modeling constraints depends quadratically on the number of memory accesses and the number of read and write ports; and linearly on the address and data widths and the number of memories. We have implemented our techniques in a prototype verification platform, and demonstrate their effectiveness on several industry designs and software programs.

**Outline** In Section 2 we give relevant background on memory semantics, SAT-based BMC, and PBA; in Section 3 we discuss the previous EMM approach; in Section 4 we describe our contributions; in Section 5 we discuss our experiments on several case studies, and conclude in Section 6.

## 2. Background

### 2.1. Bounded Model Checking (BMC)

In BMC, the specification is expressed in LTL (Linear Temporal Logic). Given a Kripke structure M, an LTL formula f, and a bound n, the translation task in BMC is to construct a propositional formula $[M, f]_n$ such that the formula is satisfiable if and only if there exists a witness of length n [16]. The satisfiability check is performed by a backend SAT solver. Verification typically proceeds by looking for witnesses or counter-examples (CE) of increasing length until the completeness bound is reached [16, 19]. The overall algorithm (BMC-1) of a SAT-based BMC method for checking a simple safety property is shown in Figure 1 (ignore lines 10-11 for now). Note that $P^i$ denotes the property node at the $i^{th}$ unrolling of the transition relation, I denotes the initial state of the system, and $LFP^i$ denotes that the path of length i is loop-free. In lines 5-7, a SAT solver is used to check the forward and backward termination criteria for correctness [19]. In line 8, a SAT solver is used to check the existence of a counter-example.

```
1.  BMC-1(n, P){ //Check safety property P within bound n
2.    CP⁰=1; LR⁻¹=Φ;
3.    for (int i=0; i<=n ; i++) {
4.      Pⁱ = Unroll(P,i); //Property node at iᵗʰ unrolling
5.      if (SAT_Solve(I∧LFPⁱ)=UNSAT or
6.         SAT_Solve(LFPⁱ∧¬Pⁱ∧CPⁱ)=UNSAT){
7.        return PROOF;} // fwd and bwd  termination check
8.      if (SAT_Solve(I∧¬Pⁱ)=SAT) return CE; //Falsify
9.      CPⁱ⁺¹ = CPⁱ∧Pⁱ;  // update CP
10.     U_Core = SAT_Get_Refutation(); // get proof of UNSAT
11.     LRⁱ = LRⁱ⁻¹ ∪ Get_Latch_Reasons(U_Core);}
12.   return NO_CE; } // No counter-example found
```

**Fig. 1. SAT-based BMC with PBA**

### 2.2. Proof-based Abstraction (PBA)

A PBA technique for SAT-based BMC is shown in lines 10-11 in Figure 1. When the SAT problem at line 8 is unsatisfiable, i.e., there is no counter-example for the safety property at a given depth i, the unsatisfiable core (U_Core) is obtained using the procedure *SAT_Get_Refutation* in line 10. This procedure simply retraces the resolution-based proof tree used by the SAT solver and identifies a subset formula that is sufficient for unsatisfiability [9, 20]. One can then use either a gate-based abstraction [9] or a latch-based abstraction [10] technique to obtain an abstract model from the *U_Core*. Here we show a latch-based abstraction technique in line 11, to obtain a set of latch reasons $LR^i$ at depth i. An abstract model is then generated for depth i by converting those latches in the given design that are *not in the set* $LR^i$ to pseudo-primary inputs. Due to the sufficiency property of *U_Core*, the resulting abstract model is guaranteed to preserve correctness of the property up to depth i [9, 10]. Depending on locality of the property, the set $LR^i$ can be significantly smaller than the total latches in the given design. Specifically in [10], a depth d (< n) is chosen such that the size of set $LR^d$ does not increase over a certain number of depths, called *stability depth*. In many cases, the property can be proved correct on the abstract model generated at depth d and hence, for the given design. One can apply PBA techniques iteratively, called iterative abstraction [10], to further reduce the set $LR^d$ and hence, obtain a smaller abstract model.

### 2.3. Memory Semantics

Embedded memories are used in several forms such as RAM, stack, and FIFO with at least one port for data access. We model a design with an embedded memory, as a *Main* module interacting with the memory module using the following memory interface signals: Address Bus (Addr), Write Data Bus (WD), Read Data Bus (RD), Write Enable (WE), and Read Enable (RE). For the single-port memory at any given clock cycle: a) at most one address is valid, b) at most one write occurs, and c) at most one read occurs. In the write phase of the memory accesses, new data is assigned to WD in the same cycle when the Addr is valid and WE is active. Note that the new written data is available for read only after the current cycle. In the read phase, data is assigned to RD in the same cycle when the Addr is valid and RE is active.

Assume that we unroll the design up to depth k (starting from 0). Let $X^j$ denote a memory interface signal variable X, at time frame j. Let the Boolean variable $E^{i,j}$ denote the address comparison between time frames i and j, defined as $E^{i,j}=(Addr^i=Addr^j)$. Then the data forwarding semantics of the memory can be expressed as follows, where j < k:

$$(E^{j,k} \wedge WE^j \wedge RE^k \wedge \forall_{j<i<k}(\neg E^{i,k} \vee \neg WE^i))$$
$$\rightarrow (RD^k = WD^j) \quad (1)$$

In the other words, the data read at depth k equals the data written at depth j if the addresses are equal at k and j, the write enable is active at j, the read enable is active at k, and for all depths strictly between j and k, no data was written at the address location $Addr^k$.

## 3. EMM for Single Memory, Read/Write Port

EMM as proposed in [18], based on the data forwarding semantics, is described as follows:
1. The MEM module is removed but the memory interface signals and their control logic are retained with their input-output directionality with respect to the *Main* Module.
2. Constraints are added at every analysis depth k in BMC, on the memory interface signals to preserve the forwarding semantics of the memory.
3. In addition, exclusivity constraints are added to improve the performance of the backend SAT solver in BMC. The idea is that when the SAT-solver decides on a valid matching read and write pair, other pairs are *implied invalid immediately*.

Note that although 1) and 2) are sufficient to generate an efficient model that preserves the validity of a correctness property, it has been shown [18] that 3) makes the performance of the SAT-based BMC superior.

The modified BMC algorithm using the EMM approach (BMC-2) for a single memory, single read/write port system (as in [18]) is shown in Figure 2. Note that the algorithm does not provide proofs with the EMM model. In this procedure, the memory modeling constraints are generated by the procedure *EMM_Constraints*, which is invoked after every unrolling. The updated constraints $C^i$ in line 5 capture the forwarding semantics of the memory up to depth i very efficiently using hybrid representations, i.e., 2-input gates and CNF clauses, in order to improve the SAT solve time. The procedure *EMM_Constraints* (lines 8-11 in Figure 2) generates the EMM constraints at a depth k by using the following 3 sub-procedures: *Generate_Addr_Equal_Sig*, *Generate_Valid_Read_Sig*, and *Generate_Read_Data_Constraints*.

```
1.  BMC-2 (n, P) {// BMC with EMM
2.   C⁻¹=φ; // initialize memory modeling constraints
3.   for (int i=0; i<=n ; i++) {
4.    Pⁱ = Unroll(P,i); // get property node at iᵗʰ unrolling
5.    Cⁱ = Cⁱ⁻¹ ∪ EMM_Constraints(i); // update the constraints
6.    if (SAT_Solve(I∧¬Pⁱ∧Cⁱ)=SAT) return CE;}
7.   return NO_CE; } // no counter-example found

8.  EMM_Constraints(k) {// Modeling of memory at depth k
9.   Generate_Addr_Equal_Sig(k);
10.  Generate_Valid_Read_Sig(k);
11.  return Generate_Read_Data_Constraints(k); }
```

**Fig. 2. SAT-based BMC with EMM**

**Generation of address comparison signals:** To capture every address pair comparison $E^{j,k}=(Addr^j=Addr^k)$, new variables $e^{j,k}_i$ and following 4 CNF clauses are added for each address bit i (where $0 \leq i < m$, and m is address width, AW)
$$(E^{j,k} \rightarrow (Addr^j_i=Addr^k_i)), ((Addr^j_i=Addr^k_i) \rightarrow e^{j,k}_i)$$
Finally, add a clause to capture the relation between $E^{j,k}$ and $e^{j,k}_i$,
$$(!e^{j,k}_0 + \ldots + !e^{j,k}_i + \ldots + !e^{j,k}_{m-1} + E^{j,k}).$$
**Generation of exclusive valid read signals:** Let the Boolean variable $s^{j,k}$ be defined as $s^{j,k}=E^{j,k} \wedge WE^j$. The decision $s^{i,k}=1$ does not necessarily imply $RD^k=WD^i$; other read-write pairs need to be established invalid through the decision procedure as well, i.e., $s^{i+1,k}=0, s^{i+2,k}=0, \ldots s^{k-1,k}=0$. Explicit constraints to capture the exclusivity of matching read and write pairs (i.e., once a matching read-write pair is chosen by the SAT-solver, the other pairs are implied invalid immediately) has been shown [18] to improve the SAT solve time significantly. Let the Boolean variables $S^{i,k}$ and $PS^{i,k}$ denote the exclusive valid read signal and the intermediate signal respectively. They are built recursively using gates for all depths i > 0 as follows:
$$\forall_{0 \leq i < k} \ PS^{i,k} = !s^{i,k} \wedge PS^{i+1,k} \quad (= RE^k \text{ for i=k})$$
$$\forall_{0 \leq i < k} \ S^{i,k} = s^{i,k} \wedge PS^{i+1,k} \quad (= PS^{0,k} \text{ for i=-1})$$
Note that $S^{i,k}=1$, immediately implies $S^{j,k}=0$ where $j \neq i$, $i,j < k$.
**Generation of constraints on read data signals:** Using the above exclusive signals, equation (1) is expressed as
$$RD^k=(S^{k-1,k} \wedge WD^{k-1}) \vee (S^{k-2,k} \wedge WD^{k-2}) \vee \ldots \vee (S^{-1,k} \wedge WD^{-1}) \quad (2)$$
where, $WD^{-1}$ denote the initial memory state. Note, for all j<k at most one $S^{j,k}$ is equal to 1. The equation (2) is expressed compactly using the following CNF clauses:
$$\forall_{0 \leq i < n}, \forall_{-1 \leq j < k} \ (S^{j,k} \rightarrow (RD^k_i = WD^j_i)) \quad (DW, \text{ data width} = n)$$
To capture validity of read signal, the following clause is added,
$$(!RE^k + S^{-1,k} + \ldots + S^{j,k} + \ldots + S^{k-1,k})$$
At depth k, the hybrid representation adds $(4 \cdot m+2 \cdot n+1) \cdot k+2 \cdot n+1$ clauses and $3 \cdot k$ gates, as compared to $(4 \cdot m+2 \cdot n+2) \cdot k+n$ gates in a purely circuit-based representation. It has also been shown that although the size of these accumulated constraints grows quadratically with depth k, they are still significantly smaller than the explicit memory-model [18].

## 4. Our Contributions

In this section, we describe our three main contributions:
1. We propose EMM for embedded systems with multiple memories, with multiple read and write ports. We show that the growth of the constraints is quadratic with analysis depth, similar to that of a single memory with a single read/write port.
2. We model arbitrary initial state of the memory precisely, and use it to provide SAT-based induction proofs in BMC.
3. We also propose combining PBA techniques with EMM. We show that using this combined approach, we can identify fewer memory modules and ports that need to be modeled; thereby reducing the model size and verification complexity.

### 4.1 EMM for Multiple Memories, Read, and Write Ports

Before we delve into a discussion of efficient modeling, we first define memory semantics in the presence of multiple read and write ports. We assume there are no data races. In other words, a memory location can be updated at any given cycle through only one write port. (We can easily extend our approach to check for data races but details are beyond the scope of the paper.) Since each memory module is accessed only through its ports, the memory modules can be considered independent of each other. In our following discussion, we first consider a single memory with multiple read and multiple write ports.

Let the design be unrolled up to depth k (starting from 0). Let $X^{j,p}$ denote a memory interface signal variable X at time frame j for a port p. Let R and W be the number of read and write ports, respectively, for the given memory. Let the Boolean variable $E^{j,i,w,r}$ denote the address comparison of the read port r at depth i, and the write port w at depth j, defined as

$E^{j,i,w,r}=(Addr^{i,r}=Addr^{j,w})$. Then the forwarding semantics of the memory can be expressed as:

$$(E^{j,k,w,r} \wedge WE^{j,w} \wedge RE^{k,r} \wedge \forall_{0 \leq p < W} \forall_{j<i<k}(\neg E^{i,k,p,r} \vee \neg WE^{i,p}))$$
$$\rightarrow (RD^{k,r} = WD^{j,w}) \quad (3)$$

In other words, data read at depth k through read port r, equals the data written at depth j through write port w, if the addresses are equal at depth k and j, write enable is active at j for the write port w, read enable is active at k for the read port r, and for all depths strictly between j and k, no data was written at the address location $Addr^{k,r}$ through any write port.

Let the Boolean variable $s^{j,k,w,r}$ be defined as $s^{j,k,w,r}=E^{j,k,w,r} \wedge WE^{j,w}$. The decision $s^{i,k,w,r}=1$ does not necessarily imply $RD^{k,r}=WD^{i,w}$; other pairs need to be established invalid through the decision procedure as well, i.e., $s^{i,k,w+1,r}=0, ..., s^{i,k,W-1,r}=0, s^{i+1,k,0,r}=0, ..., s^{i+1,k,W-1,r}=0, ..., s^{k-1,k,0,r}=0, ..., s^{k-1,k,W-1,r}=0$. Similar to the single read/write port approach [18], we add explicit constraints to capture the exclusivity of the matching read-write pair, in order to improve the SAT solve time. Let the Boolean variables $S^{i,k,w,r}$ and $PS^{i,k,w,r}$ denote the exclusive valid read signal and intermediate signal respectively for a given read port r and write port w. They are defined recursively as follows:

$PS^{k,k,0,r} = RE^{k,r}$
$\forall_{0 \leq i < k} \forall_{0 \leq p < W} PS^{i,k,p,r} = !s^{i,k,p,r} \wedge PS^{i,k,p+1,r}$ ($PS^{i,k,W,r} = PS^{i+1,k,0,r}$)
$\forall_{0 \leq i < k} \forall_{0 \leq p < W} S^{i,k,p,r} = s^{i,k,p,r} \wedge PS^{i,k,p+1,r}$ (4)

Now the forwarding semantics for multiple read and write ports can be expressed as

$$RD^{k,r} = (\vee_{0 \leq p < W, 0 \leq i < k} S^{i,k,p,r} \wedge WD^{i,p}) \vee (PS^{0,k,0,r} \wedge WD^{-1}) \quad (5)$$

Note that $S^{i,k,p,r}=1$, immediately implies $S^{j,k,q,r}=0$ where either $q \neq p$ or $j \neq i$, and $i,j < k$. Similar to [18], we use a hybrid representation to add the memory constraints as part of the procedure *EMM_Constraints*, which is invoked after every unrolling as shown in Figure 2. Given DW = n and AW = m, we give the sizes of EMM constraints added in terms of clauses and gates for each read port at a given depth k.

1. Address comparison: We require $(4 \cdot m+1) \cdot k \cdot W$ CNF clauses to represent address comparison signals.
2. Exclusive constraints: We require $3 \cdot k \cdot W$ 2-input gates to represent the exclusivity constraints in equation (4).
3. Read data constraints: We require $2 \cdot n \cdot k \cdot W+2 \cdot n+1$ CNF clauses to represent read data constraints in equation (5).

In total, we need $(4 \cdot m+2 \cdot n+1) \cdot k \cdot W+2 \cdot n+1$ clauses and $3 \cdot k \cdot W$ gates for a single read port and W write ports. For R read ports, we would need $((4 \cdot m+2 \cdot n+1) \cdot k \cdot W+2 \cdot n+1) \cdot R$ clauses and $3 \cdot k \cdot W \cdot R$ gates. Note, the growth of constraints remain quadratic with analysis depth k and is $W \cdot R$ times the constraints required for a single memory having a single read/write port. In the presence of multiple memories, we add these EMM constraints for each of them.

### 4.2. Arbitrary Initial Memory State

To model a memory with an arbitrary initial state, we introduce new symbolic variables at every time frame. Observe that for a (k-1)-depth analysis of a design, there can be at most k different memory read accesses from a single read port; out of which at most k accesses can be to un-written memory locations. Therefore, in total we need to introduce k symbolic variables for the different data words for each read port at analysis depth k-1. However, these variables are not entirely independent. Simply introducing new variables introduces additional behaviors in the verification model. Therefore, we need to identify a sufficient set of constraints that models the arbitrary initial state of the memory correctly.

Let $V^{i,p}$ and $V^{j,q}$ represent new data words introduced at depths i and j, for read ports p and q, respectively. Let $RA^{i,p}$ and $RA^{j,q}$ be the corresponding read addresses for the ports p and q (p and q need not be distinct). Let $N^{i,p}$ (and $N^{j,q}$) denote the condition that no write has occurred until depth i (and j) at address location $RA^{i,p}$ (and $RA^{j,q}$). We can then express the data read from the ports p and q at depths i and j, respectively, as:

$N^{i,p} \rightarrow (RD^{i,p}=V^{i,p})$, $N^{j,q} \rightarrow (RD^{j,q}=V^{j,q})$

Note that, if read addresses $RA^{i,p}$ and $RA^{j,q}$ are equal, then $V^{i,p}$ and $V^{j,q}$ should also be equal. We add the following constraint to capture the same,

$$(RA^{i,p}=RA^{j,q} \wedge N^{i,p} \wedge N^{j,q}) \rightarrow (V^{i,p}=V^{j,q}) \quad (6)$$

For R read ports at (k-1)-depth analysis, we need to add $k \cdot R \cdot (R-1)$ such constraints. We add these constraints using a hybrid representation in a separate sub-procedure call within the procedure *EMM_constraints*. Note that the proof step in BMC-1, (line 6, Figure 1) requires correct modeling of the arbitrary initial state of the memory. Using the new set of memory constraints as in equation (6), we augment the proof steps of BMC with EMM constraints. The modified algorithm (BMC-3) is shown in Figure 3 (ignore lines 11-12 for now). Later, we will show that the correctness of safety properties can not be shown without adding these constraints.

### 4.3. EMM with Proof-based Abstraction

As discussed earlier, EMM can significantly reduce the size of the verification model for a SoC having multiple memories and multiple ports. However, for checking the correctness of a given safety property, we may not require all the memory modules or the ports. To further reduce the model, we can abstract out *irrelevant* memory modules or ports completely. In this case, we do not need to add the memory modeling constraints for the irrelevant memory modules or ports, thereby further reducing the BMC complexity.

For the purpose of automatically identifying irrelevant memory modules and ports, we propose a technique combining EMM constraints with PBA [10]. This can not only reduce the non-memory logic (from the *Main* module) but also identify the memory modules and ports that are not required for proving correctness up to a given bounded depth of BMC analysis. The overall BMC algorithm with PBA and EMM constraints (BMC-3) is shown in lines 11-12 of Figure 3.

```
1.  BMC-3 (n, P){// Check safety property P within bound n
2.   CP^0=1; LR^{-1}=Φ; C^{-1}=Φ;
3.   for (int i=0; i<=n ; i++) {
4.    P^i = Unroll(P,i); // property node at i^{th} unrolling
5.    C^i = C^{i-1} ∪ EMM_Constraints(i); // update the constraints
6.    if (SAT_Solve(I∧LFP^i∧C^i)=UNSAT or
7.       SAT_Solve(LFP^i∧¬P^i∧CP^i∧C^i)=UNSAT) {
8.     return PROOF;} // bwd and fwd termination check
9.    if (SAT_Solve(I∧¬P^i∧C^i)=SAT) return CE;
10.   CP^{i+1} = CP^i∧P^i; // update CP
11.   U_Core = SAT_Get_Refutation(); // get proof of UNSAT
12.   LR^i = LR^{i-1} ∪ Get_Latch_Reasons(U_Core);}
13.  return NO_CE; } // no counter-example found
```

**Fig. 3. SAT-based BMC with EMM and PBA**

The dependency of the property on any memory module for a given depth i is determined easily by checking whether a latch corresponding to the control logic for that memory module (the logic driving the memory interface signals) is in the set $LR^i$. If no such latch exists in the set $LR^i$, we do not add the EMM modeling constraints for that memory module. In other words, we abstract out that memory module completely. We perform similar abstraction for each memory port. This reduces the problem size and improves the performance, as observed in our experiments reported in the next section.

## 5. Experiments

We have implemented the proposed EMM techniques in a prototype verification platform, which includes standard verification techniques for SAT-based BMC and BDD-based model checking. We report our experiences on several case studies consisting of large industry designs and software programs that have embedded memory modules with multiple read and write ports. Two case studies correspond to industry designs with many reachability properties. Another case study involves a sorting algorithm with properties validating the algorithm. For each of the properties, we require modeling of the embedded memory, and the case studies were chosen to highlight the use of our different contributions. We compare our approach (labeled EMM), with explicit memory modeling (labeled Explicit Modeling) to show the effectiveness of our approach. We experimented on a workstation with 2.8 GHz Xeon processors with 4GB running Red Hat Linux 7.2.

**Case Study on Quick Sort:** This case study makes use of EMM for multiple memories, EMM that models arbitrary initial state, and EMM with PBA to identify irrelevant memory modules.

We implemented a quick sort algorithm using Verilog HDL (Hardware Description Language). The algorithm is recursively called, first on the left partition and next on the right partition of the array (Note: a pivot partitions the array into left and right). We implemented the array as a memory module with AW=10 and DW=32, with 1 read and 1 write port. We implemented the stack (for recursive function calls) also as a memory module with AW=10 and DW=24, with 1 read and 1 write port. The design has 200 latches (excluding memory registers), 56 inputs, and ~9K 2-input gates. We chose two properties: a) P1: the first element of the sorted array (in ascending order) can not be greater than the second element, b) P2: after return from a recursive call, the program counter should go next to a recursive call on the right partition or return to the parent on the recursion stack. The array is allowed to have arbitrary values to begin with. This requires precise handling of the arbitrary initial memory state (equation (6)) to show the correctness of the property.

For different array sizes N, we compared the performance of EMM and Explicit Modeling approaches, using the forward induction proof checks in BMC-3 and BMC-1 respectively. We used a time limit of 3 hours for each run. We present the results in Table 1. Column 1 shows different array sizes N; Column 2 shows the properties; Column 3 shows the forward proof diameter; Columns 4-5 and 6-7 show performance time and space used by EMM and Explicit Modeling, respectively. Note that using EMM we were able to prove all properties in the given time limit, while Explicit Modeling simply times out.

**Table 1. Performance summary on Quick Sort**

| N | Prop | D | EMM | | Explicit | |
|---|---|---|---|---|---|---|
| | | | Sec | MB | Sec | MB |
| 3 | P1 | 27 | 64 | 55 | >3hr | NA |
| 3 | P2 | 27 | 30 | 44 | >3hr | NA |
| 4 | P1 | 42 | 601 | 105 | >3hr | NA |
| 4 | P2 | 42 | 453 | 124 | >3hr | NA |
| 5 | P1 | 59 | 6376 | 423 | >3hr | NA |
| 5 | P2 | 59 | 4916 | 411 | >3ht | NA |

Note that property P1 depends on both the array and the stack, while property P2 depends on only the stack for correctness. In other words, for P2, the contents of the array should not matter at all. We used the PBA technique to examine this. For property P2, we compared performance of EMM with PBA using BMC-3, with that of PBA on Explicit Modeling using BMC-1. We used a stability depth of 10 to obtain the stable set LR. We present the results in Table 2. Column 1 shows different array sizes N, Columns 2-5 show performance figures for EMM. Specifically, Column 2 shows the number of latches in the reduced model size using EMM with PBA. The value in bracket shows the original number of latches. Column 3 shows the time taken (in sec) for PBA to generate a stable latch set. Columns 4-5 show the time and memory required for EMM to provide the forward induction proof. Columns 6-9 report these performance numbers for the Explicit Modeling.

It is interesting to note that by use of PBA, the reduced model in Column 2 did not have any latch from the control logic of the memory module representing the array. Therefore, we were able to automatically abstract out the entire array memory module, while doing BMC analysis on the reduced model using EMM. Note that this results in significant improvement in performance, as clear from a comparison of the performance figures of EMM on property P2 in columns 4-5 of Tables 1 and 2. Moreover, we see several orders of magnitude performance improvement over the Explicit Modeling, even on the reduced models. Note, for N=5 we could not generate a stable latch model in the given time limit for the Explicit Modeling case.

**Table 2. Performance summary on Quick Sort on P2**

| N | EMM +PBA | | EMM-Proof on Red. Model | | Explicit+PBA | | Explicit on Red. Model | |
|---|---|---|---|---|---|---|---|---|
| | FF (orig) | Sec | Sec | MB | FF (orig) | Sec | Sec | MB |
| 3 | 91 (167) | 10 | 5 | 13 | 293 (37K) | 293 | 2K | 274 |
| 4 | 93 (167) | 38 | 145 | 40 | 2858 (37K) | 2858 | 10K | 456 |
| 5 | 91 (167) | 351 | 2316 | 116 | - (37K) | >3hr | NA | NA |

**Case Study on Industry Design I:** This case study makes use of our approach of EMM for multiple memories and EMM with induction proofs.

The industry design is a low-pass image filter with 756 latches (excluding the memory registers), 28 inputs and ~15K 2-input gates. It has two memory modules, both having address width, AW = 10 and data width, DW = 8. Each module has 1 write and 1 read port, with memory state initialized to 0. There are 216 reachability properties.

EMM: We were able to find witnesses for 206 of the 216 properties, in about 400s requiring 50Mb. The maximum depth over all witnesses was 51. For the remaining 10 properties, we were able to obtain the proofs by induction using BMC-3, in less than 1s requiring 6Mb. Note that the introduction of new variables to model arbitrary initial memory state, without the constraints in equation (6), was sufficient for the proofs although they capture extra behavior in the verification model.

Explicit Modeling: We required 20540s (~6Hrs) and 912Mb to find witnesses for all 206 properties. For the remaining 10 properties, we were able to obtain the proofs by induction using BMC-1 in 25s requiring 50Mb.

**Case Study on Industry Design II:** This case study makes use of EMM for memory with multiple ports, and for finding invariants that can aid proofs by induction.

The design has 2400 latches (excluding the memory registers), 103 inputs and ~46K 2-input gates. It has one memory module with AW=12 and DW=32. The memory module has 1 write port and 3 read ports, with memory state initialized to 0. There are 8 reachability properties.

We found spurious witnesses at depth 7 for all properties, if we abstract out the memory completely. Thus, we needed to include the memory module. Using EMM, we were not able to find any witnesses for these properties up to depths of 200 in about 10s. Next, we tried obtaining a proof of unreachability for all depths. Using EMM with PBA, we were able to reduce the model to about 100 latches requiring 4-5 minutes. However, the model was not small enough for our BDD-based model checker or SAT-based BMC to provide a proof. We also noticed that the WE (write enable) control signal stayed inactive in the forward search of 200 depth. Observing that, we hypothesized that the memory state does not get updated, i.e., it remains in its initial state. This is expressed using the following LTL property:

$$G(WE=0 \text{ or } WD=0)$$

i.e., always, either the write enable is inactive or the write data (WD) is 0. Using BMC-3, we were able to prove the above property using backward induction at depth 2 in less than 1s. Explicit Modeling using BMC-1 takes 78s to prove the same.

The above invariant implies that the data read is always 0 (could potentially be a design bug). Next we abstracted out the memory, but applied this constraint to the input read data signals. We used PBA to further reduce the design to only 20-30s latches for each property (taking about a minute). We then proved each property unreachable on the reduced model using forward induction proof in BMC-1 in less than 1s. (Our BDD-based model checker was unable to build even the transition relation for these abstract models.)

## 6. Conclusions

We have proposed several techniques for verifying embedded memory systems using EMM. We extend the previous EMM approach for a single memory with a single read/write port, to the more commonly occurring memory systems of multiple memories with multiple read and write ports. We also extend the previous EMM approach for falsification to derivation of proofs. We have proposed a precise modeling of the arbitrary initial state of memory, for use in SAT-based induction proofs using BMC. We have also proposed combining PBA techniques with EMM. We showed that using this combined approach, we can identify fewer memory modules that need to be modeled; thereby reducing the model size and verification problem complexity. We applied these EMM techniques on several case studies to show their effectiveness in practice, in comparison to an explicit memory modeling approach. In one case study, EMM techniques also helped to efficiently check invariants, which were then used to prove several properties unreachable.

## References

[1] M. Pandey, R. Raimi, D. L. Beatty, and R. Brayton, "Formal Verification of PowerPC Arrays using Symbolic Trajectory Evaluation," *Proceedings of DAC*, 1996.
[2] C. J. H. Seger and R. E. Bryant, "Formal Verification by Symbolic Evaluation by Partially-ordered Trajectories," *Proceedings of Formal Method in System Design*, 1995.
[3] K. L. McMillan, *Symbolic Model Checking: An Approach to the State Explosion Problem*: Kluwer Academic Publishers, 1993.
[4] E. M. Clarke, O. Grumberg, and D. Peled, *Model Checking*: MIT Press, 1999.
[5] D. E. Long, "Model checking, abstraction and compositional verification," Carnegie Mellon University, 1993.
[6] E. M. Clarke, O. Grumberg, S. Jha, Y. Lu, and H. Veith, "Counterexample-guided abstraction refinement," in *Proceedings of CAV*, vol. 1855, *LNCS*, 2000, pp. 154-169.
[7] E. M. Clarke, A. Gupta, J. Kukula, and O. Strichman, "SAT based abstraction-refinement using ILP and machine learning techniques," in *Proceedings of CAV*, 2002.
[8] P. Chauhan, E. M. Clarke, J. Kukula, S. Sapra, H. Veith, and D. Wang, "Automated Abstraction Refinement for Model Checking Large State Spaces using SAT based Conflict Analysis," in *Proceedings of FMCAD*, 2002.
[9] K. McMillan and N. Amla, "Automatic Abstraction without Counterexamples," in *Proceedings of TACAS*, April 2003.
[10] A. Gupta, M. Ganai, P. Ashar, and Z. Yang, "Iterative Abstraction using SAT-based BMC with Proof Analysis," in *Proceedings of ICCAD*, 2003.
[11] J. R. Burch and D. L. Dill, "Automatic verification of pipelined microprocessor control," in *Proceedings of CAV*, 1994.
[12] M. N. Velev, R. E. Bryant, and A. Jain, "Efficient Modeling of Memory Arrays in Symbolic Simulation," in *Proceedings of CAV*, 1997.
[13] R. E. Bryant, S. German, and M. N. Velev, "Processor Verification Using Efficient Reductions of the Logic of Uninterpreted Functions to Propositional Logic," in *Proceedings of CAV*, 1999.
[14] M. N. Velev, "Automatic Abstraction of Memories in the Formal Verification of Superscalar Microprocessors," in *Proceedings of TACAS*, 2001.
[15] R. E. Bryant, S. K. Lahiri, and S. A. Seshia, "Modeling and Verifying Systems using a Logic of Counter Arithmetic with Lambda Expressions and Uninterpreted Functions," in *Proceedings of CAV*, 2002.
[16] A. Biere, A. Cimatti, E. M. Clarke, and Y. Zhu, "Symbolic Model Checking without BDDs," in *Proceedings of TACAS*, 1999.
[17] M. Ganai, A. Gupta, and P. Ashar, "Distributed SAT and Distributed Bounded Model Checking," in *Proceedings of CHARME*, 2003.
[18] M. Ganai, A. Gupta, and P. Ashar, "Efficient Modeling of Embedded Memories in Bounded Model Checking," in *Proceedings of CAV*, 2004.
[19] M. Sheeran, S. Singh, and G. Stalmarck, "Checking Safety Properties using Induction and a SAT Solver," in *Proceedings of FMCAD*, 2000.
[20] L. Zhang and S. Malik, "Validating SAT Solvers Using an Independent Resolution-Based Checker: Practical Implementations and Other Applications," in *Proceedings of DATE*, 2003.
[21] M. Ganai, L. Zhang, P. Ashar, and A. Gupta, "Combining Strengths of Circuit-based and CNF-based Algorithms for a High Performance SAT Solver," in *Proceedings of DAC*, 2002.